# Mining Scientific Papers for Bibliometrics: a (very) Brief Survey of Methods and Tools

Iana Atanassova[1], Marc Bertin[2] and Philipp Mayr[3]

[1] *iana.atanassova@univ-fcomte.fr*
*Centre Tesniere, University of Franche-Comte, France*

[2] *bertin.marc@gmail.com*
Centre Interuniversitaire de Rercherche sur la Science et la Technologie (CIRST),
Université du Québec à Montréal (UQAM), Canada

[3] *philipp.mayr@gesis.org*
GESIS, Leibniz Institute for the Social Sciences, Germany

**Introduction**

The Open Access movement in scientific publishing and search engines like Google Scholar have made scientific articles more broadly accessible. During the last decade, the availability of scientific papers in full text has become more and more widespread thanks to the growing number of publications on online platforms such as ArXiv and CiteSeer (Wu, 2014). The efforts to provide articles in machine-readable formats and the rise of Open Access publishing have resulted in a number of standardized formats for scientific papers (such as NLM-JATS, TEI, DocBook).

*Corpora*

Different projects have been carried out to respond to the need of full-text datasets for research experiments (PubMed, JSTOR, etc.) and corpora.
E.g. the *iSearch* dataset was designed to facilitate research and experimentation in information retrieval, and specifically in aspects of task-based and integrated (a.k.a. aggregated) search. Its compressed size is about 46GB of documents in English from the physics domain that were collected from public libraries and open archive resources.

*Semantic Web and Information Retrieval*

Scientific papers are highly structured texts and display specific properties related to their references but also argumentative and rhetorical structure. Recent research in this field has concentrated on the construction of ontologies for citations and scientific articles.
CiTO (Shotton, 2010), the Citation Typing Ontology, is an ontology for the characterization of citations, both factually and rhetorically. It is part of SPAR, a suite of Semantic Publishing and Referencing Ontologies. Other SPAR ontologies are described at http://purl.org/spar/.

**Statistical Analysis of Textual Data**

*Text Mining in R*

Temis, an R Commander plugin (Bastin, 2013) provides integrated tools for text mining. Corpora can be imported in raw text. Another package is IRaMuTeQ (Ratinaud, 2009), a python application which uses the R libraries.

*Correspondence Analysis*

Correspondence analysis is a technical description of contingency tables and is mainly used in the field of text mining (Morin, 2006).
These tools could be very useful on the perspectives for the development of new text analytics approaches for bibliometrics.

**Natural Language Processing Tools**

Research in the field of Natural Language Processing (NLP) has provided a number of open source tools for versatile text processing.
The Apache *OpenNLP* library (Baldridge, 2005) is a machine learning based toolkit for the processing of natural language text. Written in Java, it is open source and platform-independent.
Stanford *CoreNLP* (Manning, 2014) integrates many NLP tools, including a part-of-speech (POS) tagger, a named entity recognizer (NER), a parser, a coreference resolution system, a sentiment analysis tool, and bootstrapped pattern learning tools. Stanford CoreNLP is written in Java and licensed under the GNU General Public License
*MALLET* (McCallum, 2002) is a Java-based package for statistical NLP, document classification, clustering, topic modeling, information extraction, and other machine learning applications to text. It includes sophisticated tools for document classification: efficient routines for converting text to "features", a wide variety of algorithms (including Naïve Bayes, Maximum Entropy, and Decision Trees), and code for evaluating classifier performance using several common metrics.

*GATE* (Cunningham, 2002) is open source free software for all types of computational tasks involving human language. It includes components for diverse NLP tasks, e.g. parsers, morphology, tagging, Information Retrieval tools, Information Extraction components for various languages.

*CiteSpace* (Chen, 2006) is a freely available Java application for visualizing and analyzing trends and patterns in scientific literature. It is designed to answer questions about a knowledge domain, which is a broadly defined concept that covers a scientific field, a research area, or a scientific discipline.

**What is next?**

Several studies examine the distribution of references in papers (Bertin, 2013). However, up to now full-text mining efforts are rarely used to provide data for bibliometric analyses. An example is the special issue on Combining Bibliometrics and Information Retrieval (Mayr, 2015). Novel approaches to full-text processing of scientific papers and linguistic analyses for Bibliometrics can provide insights into scientific writing and bring new perspectives to understand both the nature of citations and the nature of scientific articles. The possibility to enrich metadata by the full-text processing of papers offers new fields of application to bibliometrics studies like e.g. text reuse patterns in specific disciplines.

Working with full text allows us to go beyond metadata used in Bibliometrics. Full text offers a new field of investigation, where the major problems arise around the organization and structure of text, the extraction of information and its representation on the level of metadata. Unlike text-mining from titles and abstracts, full-text processing allows the extraction of rhetorical elements of scientific discourse, such as results, methodological descriptions, negative citations, discussions, etc. Scientific abstracts, by summarizing the text, provide only short, synthetic and thematic information.

Furthermore, the study of contexts around in-text citations offers new perspectives related to the semantic dimension of citations. The analyses of citation contexts and the semantic categorization of publications will allow us to rethink co-citation networks, bibliographic coupling and other bibliometric techniques.

Our aim is to stimulate research at the intersection of Bibliometrics and Computational Linguistics in order to study the ways Bibliometrics can benefit from large-scale text analytics and sense mining of scientific papers, thus exploring the interdisciplinarity of Bibliometrics and Natural Language Processing. Typical questions of this emerging field are: How can we enhance author network analysis and Bibliometrics using data obtained by text analytics? What insights can NLP provide on the structure of scientific writing, on citation networks, and on in-text citation analysis?